\documentclass[
 reprint,
 amsmath,
 amssymb,
 aps
]{revtex4-2}

\usepackage{graphicx}
\usepackage{hyperref}
\usepackage[linesnumbered, ruled, vlined, titlenotnumbered, noend]{algorithm2e}
\usepackage{amsmath, amsfonts, amssymb, amsthm, color}
\usepackage{tikz-cd}
\usepackage[capitalise]{cleveref}
\crefname{algocf}{alg.}{algs.}
\Crefname{algocf}{Algorithm}{Algorithms}

\newcommand{\ket}[1]{|#1\rangle}

\newcommand{\expnumber}[2]{{#1}\mathrm{E}{#2}}
\newcommand{\UER}[1]{\mathrm{UER}_{\braket{#1}}}
\newcommand{\LER}[1]{\mathrm{LER}_{\braket{#1}}}
\newcommand{\CE}[1]{\mathrm{CE}_{\braket{#1}}}
\newcommand{\wt}[1]{\mathrm{wt}({\mathbf{#1}})}
\newcommand{\subtext}[2]{{#1}_\mathrm{#2}}
\usepackage{braket}
\usepackage{makecell}
\usepackage{subcaption}

\newcommand{\added}[1]{{#1}}

\begin{document}

\author{Mark Webster}
\affiliation{IonQ Inc.}
\thanks{\ Corresponding author: \url{mark.webster@ionq.co}}
\author{Nicolas Delfosse}
\affiliation{IonQ Inc.}

\title{\added{Efficient} logical operations in quantum LDPC codes using simple resource states}

\date{\today}

\begin{abstract}
Quantum LDPC codes provide a substantial reduction in qubit overhead required for fault-tolerant quantum computation compared to surface code, thanks to their high encoding rate. However, operating simultaneously on multiple logical qubits encoded in the same block is more challenging and may slow down logical operations.
Prior work addresses this problem by designing complex resource states to perform logical measurements in LDPC codes.
Here, we propose an approach that only consumes cat states. Whereas previous work on cat-based measurements focuses on a single logical measurement, we design a protocol for the joint measurement of $\ell$ commuting logical operators.
The key ingredient is the design of a scheduler code determining the measurement sequence and allowing for the decoding of all logical measurement outcomes.
Numerical simulations with the LDPC codes Q70 and Q102 of the walking cat architecture show a \added{reduction of nearly $3\times$ in the number of cat-based measurements required versus the Viterbi method for measuring} $\ell=20$ commuting logical operators.
Combining our logical measurement protocols with a new variant of the CliNR partial error correction scheme, we achieve a \added{reduction of up to $73\times$ versus the Viterbi method} for random Clifford circuits. Our approach also applies to non-Clifford gates, producing a \added{reduction of up to $5\times$ versus the Viterbi method} for Toffoli gates.

\end{abstract}

\maketitle
 
\section{Introduction}\label{sec:introduction}

A quantum error correction code lies at the core of any fault-tolerant quantum computer (FTQC) architecture.
Motivated by theoretical breakthroughs~\cite{hastings2021fiber, panteleev2022asymptotically, leverrier2022quantum, breuckmann2021quantum}, promising numerical results~\cite{tremblay2022constant, bravyi2024high, scruby2026high, pecorari2025high, aydin2025cyclic, zhao2026towards}, architecture studies~\cite{xu2024constant, ruiz2025ldpc, yoder2025tour, webster2026pinnacle, cain2026shor, khan2026architecting, walking_cat}, and experimental realizations~\cite{wang2026demonstration, tham2026breakeven}, quantum low-density parity-check (LDPC) codes are gaining traction.
In contrast to surface codes, they encode many logical qubits in the same block, reducing the qubit-overhead.
However, this encoding makes logical qubits harder to access because operating simultaneously on logical qubits supported on the same block is challenging, resulting in a potential slowdown of logical operations.

Substantial effort has been invested in the design of fault-tolerant logical measurements for quantum LDPC codes based on graph-based resource states~\cite{cohen_ldpc, huang2023homomorphic, cowtan2024css, swaroop2024universal, cross2024improved, cowtan2024ssip, he2025extractors, baspin2025fast, yuan2026parsimonious, chang2026constant}.
This approach has three main limitations
(a) It relies on complex resource states and for each logical measurement, a new code is formed, merging the memory with the resource state, imposing stringent hardware requirements to accommodate all these merged codes.
(b) These protocols are typically described in terms of homological algebra which is a beautiful formalism but may be less accessible for part of the community unfamiliar with this framework.
(c) Finally, these protocols are mostly studied through asymptotic proofs of fault-tolerance and their practical performance needs further investigation.

To remove the need for complex resource states, we leverage Shor's measurement scheme which only consumes cat states $\ket{0\dots 0} + \ket{1\dots 1}$~\cite{shor1996fault}.
Each cat state interacts with the memory only for a single time step, before it is measured, removing the need to form a merge code.
The exact protocol is easy to understand by inspecting the circuit without additional formalism.
Furthermore, we focus on the practical regime and we leverage the fact that fault-tolerance is not required to achieve the target logical error rate required for a specific quantum algorithm.
Our approach requires bringing cat states next to the qubits to measure, making this protocol well-suited for moving qubits such as trapped ions~\cite{kielpinski2002architecture}, neutral atoms~\cite{bluvstein2024logical}, spin qubits~\cite{loss1998quantum} or electrons floating on helium~\cite{castoria2026selective}.

Shor's method was originally proposed for the measurement of syndrome bits and improved in~\cite{delfosse2020shortshorstylesyndromesequences, Tansuwannont_2023,wang2026adaptivelosstolerantsyndromemeasurements}.
The walking cat architecture demonstrates the practicality of cat-based measurements for logical measurements in quantum LDPC codes~\cite{walking_cat}.
Therein, the measurement of a logical operator is performed through a Viterbi measurement which reaches a target accuracy.

In this work, we design substantially \added{more efficient} cat-based logical measurement protocols by combining the measurement of $\ell$ commuting logical operators.
Our main innovation is the introduction of a scheduler code that determines the sequence of cat-based measurements and enables the joint decoding of all the measurement outcomes.
To illustrate our results, we consider the quantum LDPC codes Q70 and Q102 from the walking cat architecture.
The measurement of $\ell=10$ and $20$ commuting logical operators consumes respectively an average of only $2.1$ and $1.7$ cat states per logical measurement.
Our results also apply to trivial logical measurement, that is stabilizers, which leads to a protocol for  preparation of a stabilizer state through the measurement of its stabilizer generators.

We introduce a new variant of the CliNR partial error correction  scheme~\cite{CliNR_original, CliNR_optimised} which leverages \added{our logical measurement protocols to improve the efficiency of logical Toffoli gates and logical Clifford gates}.
For Q70 and Q102, we obtain  a $4.2 \times$ and $5.1 \times$ \added{reduction respectively in the number of cat-based measurements required versus the Viterbi method} for Toffoli gates and a $18.6\times$ and $73.1\times$ \added{reduction} for random Clifford operations.
This application of CliNR is unexpected because CliNR was originally designed for partial error correction but the original scheme leads to an increase in execution time.
To achieve these results, we execute part of the scheme at the logical level and other parts at the physical level. 
Moreover, we leverage \added{our logical measurement protocols to improve the efficiency of resource state preparation and verification for the CliNR scheme}. 

\added{\section{Previous Work on Measurement Outcomes and Classical Codes}
The idea of protecting measurement outcomes with a classical code has a long history. 
Fujiwara first observed that redundant stabilizer measurements can be used to protect syndrome data against measurement errors \cite{PhysRevA.90.062304}. 
This led to the framework of data-syndrome codes, where data and syndrome errors are corrected jointly \cite{6874892,DBLP:conf/isit/AshikhminLB16}. 
It was later connected to the single-shot properties of quantum codes \cite{Campbell_2019,Delfosse_2022}. 

The same principle has been applied to logical measurements. 
Sequences of logical measurements encoded with a classical code were discussed in \cite{delfosse2020shortshorstylesyndromesequences} for small block codes. 
In \cite{PRXQuantum.3.010331} temporally encoded lattice surgery (TELS) measures a set of commuting logical Pauli operators, together with redundant products of them chosen from a classical code. 
This allows timelike failures in surface-code lattice surgery to be detected, so fewer syndrome rounds are needed per lattice-surgery operation. 
More recently, \cite{wills2026concatenatingalgebraiccodeshighrate} used timelike Reed–Solomon protection against measurement errors in a Shor-style syndrome extraction scheme for Galois qudits concatenated over high-rate LDPC codes. 

Our protocol differs from these works in three ways. First, the scheduler code can be used for error correction as well as detection: the decoder recovers all $\ell$ logical measurement outcomes, rather than flagging a failure and repeating the measurement. 
Secondly, we target quantum LDPC codes with many logical qubits per block using cat states for logical measurements and use simulation results for these codes indicating that measurement outcomes suffer from independent bit flip errors \cite{walking_cat}. 
Thirdly, we give an algorithm for choosing the scheduler code to minimize the number of cat-based measurements required to meet a desired logical error rate.}

\section{Background: Measurement of a single logical operator}\label{sec:meas_intro}
Our primitive operation is a {\em cat-based measurement} which performs the measurement of a $\overline{w}$-qubit Pauli operator using a $\overline{w}$-qubit cat state.
Following~\cite{walking_cat}, we assume that a cat-based measurement can be executed at each syndrome  extraction cycle (SEC) and that measurement outcomes suffer from independent bit-flips with flip rate $p_F$.
An optimized cat state preparation scheme is described in~\cite{walking_cat}, which also verifies numerically the independence of cat-based measurement flips. 
We expect this assumption to hold for quantum LDPC codes with single-shot properties~\cite{gu2024single}.

A \textit{measurement protocol} takes as input a set of $\ell$ \textit{target Pauli operators} to measure.
It specifies a \textit{schedule} of $m$ measurements which may be a product of the target Paulis and which have sufficient redundancy to cater for measurement flip errors.
Measurement flip errors are represented by a vector $\mathbf{e} \in \{0,1\}^m$ with distribution $P(\mathbf{e}) = p_F^{\wt{e}}(1-p_F)^{m-\wt{e}}$.
The physical measurement outcomes resulting from the schedule produce a binary \textit{outcome vector} $\mathbf{v}$ of length $m$.
The outcome vector is used to calculate the most likely \textit{logical outcome vector} $\hat{\mathbf{u}}$ of length $\ell$.
The protocol must include enough measurements to ensure that the probability of any logical measurement being incorrect is less than the \textit{target logical error rate} $\varepsilon$.

In \cite{walking_cat} three protocols for measurement of a single logical Pauli are introduced which we extend to sets of multiple commuting Paulis in this work. 
The first is the \textit{error detected measurement} protocol (EDM) where the target Pauli is measured $m$ times in each attempt. 
If all measurement outcomes are the same, the protocol succeeds. Otherwise, the protocol is restarted. 
Providing $m>\log\varepsilon/\log p_F$, the target logical error rate is achieved.
In the \textit{error corrected measurement} protocol (ECM), a single attempt of $m$ measurements is conducted and the most likely outcome is selected by majority vote. For this protocol to meet the target logical error rate, $m$ must satisfy $\sum_{\wt{e}\le(m-1)/2}P(\mathbf{e}) > 1-\varepsilon$.
The \textit{Viterbi measurement protocol} is similar to the ECM protocol but terminates as soon as $|m-2m_1|>\log(\varepsilon^{-1}-1)/\log(p_F^{-1}-1)$ where $m_1$ is the weight of the outcome vector $\mathbf{v}$.
We use the Viterbi measurement protocol as the base case for comparison with our new multi-Pauli measurement protocols as it requires the lowest number of cat-based measurements on average of these methods.

\section{Measurement of Commuting Paulis}\label{sec:fast_measurements}
In this section we introduce the new MEDM and MECM protocols for measuring sets of commuting Pauli operators.
The new protocols require fewer cat-based measurements per Pauli operator than the protocols of \cite{walking_cat}.

\subsection{Scheduler Matrices and Codes}\label{sec:scheduler_codes}
In the MEDM and MECM protocols we measure $\ell$ independent commuting Pauli operators $P_0,\dots,P_{\ell - 1}$  using a measurement schedule  specified by  an $\ell\times m$ generator matrix $G$. 
We refer to $G$ as the \textit{scheduler matrix}, and to the binary linear code generated by $G$ as the \textit{scheduler code}, denoted $\braket{G}$.
The number of columns in $G$, $|G| = m$, is the number of measurements in the schedule. 
Each row of $G$ corresponds to one of the target Paulis to be measured. 
Each column of $G$ corresponds to a product of the target Pauli operators so that the $j$th measurement is $Q_j :=\prod_{0 \le i < \ell} P_i^{G_{ij}}$.
The measurement outcome vector $\mathbf{v}$ is of length $m$ and the $j$th entry corresponds to the measurement outcome of $Q_j$.
We require that the first $\ell$ columns of $G$ form an \textit{information set} ({\em i.e.,} a maximum set of independent columns of $G$) which allows us to infer the logical measurement outcome $\mathbf{u}$ in the absence of error such that $\mathbf{v}=\mathbf{u}G$.
If $G$ is in systematic form $G = [I|A]$, $\mathbf{u}$ is determined by taking the first $\ell$ bits of $\mathbf{v}$.

\subsection{MEDM Protocol}\label{sec:MEDM}
We generalize the EDM protocol to a set of $\ell$ commuting Paulis by using an $\ell \times m$ scheduler matrix $G$.
After $m$ measurements, we use the rank $m-\ell$ check matrix $H$ satisfying $GH^T=0$ to detect errors. 
If $\mathbf{v}H^T \ne 0$ the MEDM protocol restarts.
Otherwise, no error is detected, and we return $\mathbf{u}$ by examining the first $\ell$ bits of $\mathbf{v}$.

The \textit{undetectable error rate} sums the probabilities of the non-zero codewords of $\braket{G}$ so that $\UER{G} := \sum_{\mathbf{0 \ne e} \in \braket{G}}P(\mathbf{e})$ and is useful for analyzing the performance of the MEDM protocol.
For a given attempt, the MEDM protocol terminates if there is either no error or the error is a non-zero codeword, and so the average number of attempts is given by:
\begin{align}
\subtext{N}{Avg} := 1/(P(\mathbf{0}) + \UER{G}).\label{eq:Ravg}
\end{align}

Once MEDM terminates, there is a logical error only if the error is a non-zero codeword in $\braket{G}$ and the target logical error rate is attained providing:
\begin{align}
    U_{\braket{G}} :=\UER{G}/(P(\mathbf{0}) + \UER{G})< \varepsilon.\label{eq:UER_requirement}
\end{align}

We can reduce the average number of measurements per attempt by terminating the MEDM protocol as soon as an error is detected and we refer to this method as the \textit{truncated MEDM protocol}. 
Once $\ell$  measurements have been completed, the error-free logical measurement outcome $\mathbf{u}$ is determined, and we terminate if measurement outcomes are different to the error-free codeword $\mathbf{u}G$. 
Based on the tail-sum formula, the average number of measurements per attempt depends on the UERs of the codes formed from the first $i$ columns of $G$ (denoted $G|_i$) and is given by:
\begin{align}
    m_N := \ell + 1 + \sum_{\ell+1 \le i \le m-1}\left((1-p_F)^i+\UER{G|i}\right)\label{eq:mR}
\end{align}
The average number of cat-based measurements is $\subtext{m}{Avg} = m\subtext{N}{Avg}$ for the full MEDM protocol and $\subtext{m}{Avg} = m_N\subtext{N}{Avg}$ for the truncated MEDM protocol.

\subsection{MECM Protocol}\label{sec:MECM}
We now explain how to generalize the ECM protocol of \cite{walking_cat} to $\ell$ commuting Paulis using an $\ell \times m$ scheduler matrix $G$.
A decoder $\mathcal{D}$ is applied to the measurement outcome vector $\mathbf{v}$ to find the most likely logical outcome $\hat{\mathbf{u}}$.
The most likely logical outcome is the $\hat{\mathbf{u}}$ which minimizes the weight of the \textit{correction} $\hat{\mathbf{e}} = \mathbf{v} + \hat{\mathbf{u}}G$.
Each decoder has a set of \textit{correctable errors} $\CE{G} = \{\mathbf{e} : \mathcal{D}(\mathbf{e} + \mathbf{u}G) = \mathbf{u}, \forall \mathbf{u}\}$.
The \textit{logical error rate} of $\braket{G}$ is $\LER{G} = 1-P(\CE{G})$.
Providing the scheduler code satisfies $\LER{G} < \varepsilon$, the target logical error rate is achieved.

For small $\ell,m$ we use a look up decoder which has ideal performance.
For large $m$ we employ heuristic decoders such as information set decoders \cite{prange_ISD,Lee1988AnOO, stern_ISD}, trellis \cite{viterbi} or belief propagation decoders \cite{belief_propagation} (where the code is LDPC or LDGM \cite{LDPC}).
We describe a random information set decoder in more detail in \Cref{sec:WE_heuristic}. 
For large $m$ we estimate the logical error rate by assuming  all errors of weight $(d-1)/2$ are correctable where $d$ is the distance of the code:
\begin{align}
    \widehat{\text{LER}}_{\braket{G}} =1- \sum_{\wt{e}\le (d-1)/2}P(\mathbf{e}) \cdot\label{eq:approx_LER}
\end{align}

We generalize the Viterbi measurement protocol of \cite{walking_cat} to multiple Paulis by terminating MECM once the \textit{posterior probability} $P(\hat{\mathbf{u}}|\mathbf{v}) > 1-\varepsilon$.
This reduces the average number of measurements required and we refer to this method as the \textit{truncated MECM protocol}.
We explain how to calculate the posterior probability exactly for small $\ell,m$ in \Cref{sec:WE+UER} and using random information set sampling for larger $\ell,m$ in \Cref{sec:WE_heuristic}.

\subsection{Choice of Scheduler Code for Truncated MECM Protocol}\label{sec:scheduler_choice}
In this section we show how to choose a scheduler code which minimizes the average number of cat-based measurements required by the truncated MECM protocol. 
Let the number of cat-based measurements required for the error $\mathbf{e}$ in the truncated MECM protocol be $m_\mathbf{e}$.
The average number of cat-based measurements for the truncated MECM protocol is $\subtext{m}{Avg} = \sum_\mathbf{e}m_eP(\mathbf{e})$.
Where $mp_F\ll 1$, the highest contributor to $\subtext{m}{Avg}$ is the number of cat-based measurements $m_0$ required for the all-zero error string. 
For the all-zero error, the measurement outcome vector is a codeword of form  $\mathbf{v} = \mathbf{u}G$.
In this case, the protocol terminates once the posterior probability $1-P(\mathbf{u|v}) = \UER{G}/(P(\mathbf{0}) + \UER{G}) < \varepsilon$ which is the same condition as in \Cref{eq:UER_requirement}.

To minimize the average number of cat-based measurements for the truncated MECM protocol, we use a scheduler matrix of form $G = [G_0|G_1]$.
For $G_0$ we choose the code with the smallest block size meeting the UER condition in \Cref{eq:UER_requirement}.
This guarantees that the truncated MECM protocol terminates at the last column of $G_0$ for $\mathbf{e=0}$ and so $m_0=|G_0|$ and is minimal.
We then choose the smallest $G_1$ that has distance sufficiently large to guarantee that $G:=[G_0|G_1]$ satisfies $\LER{G}<\varepsilon$. 
For our simulation, we choose codes $G_0$ and $G_1$ from the best-known-distance binary linear codes of \cite{codetables}.
In \Cref{sec:MEDM_MECM_accessibility} we show how to generate scheduler matrices to accommodate restrictions on the accessibility of Pauli measurements in the walking cat architecture (see \Cref{sec:cliffords_walking_cat}).

\section{Logical Clifford and Toffoli Gates}\label{sec:fast_clifford_ops}
\begin{figure}[hbt]
    \centering
    \begin{subfigure}[t]{0.23\textwidth}
        \centering
        \includegraphics[width=\textwidth]{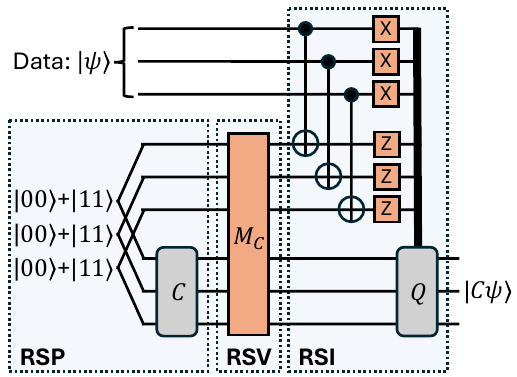}
        \caption{Physical CliNR}\label{fig:physical_CliNR}
    \end{subfigure}%
    ~ 
    \begin{subfigure}[t]{0.23\textwidth}
        \centering
        \includegraphics[width=\textwidth]{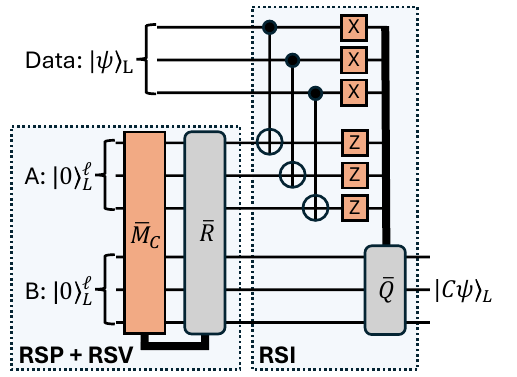}
        \caption{Logical CliNR}\label{fig:logical_CliNR}
    \end{subfigure}
    \caption{Physical and Logical CliNR Circuits: in physical CliNR, resource state preparation (RSP) and verification (RSV) are separate steps followed by injection (RSI).
    In logical CliNR, RSP and RSV are merged and implemented at the logical level and RSI is implemented at the physical level.}
    \label{fig:CliNR_circuit}
\end{figure}

Here, we propose a new protocol for logical Clifford gates, and we apply it to build more efficent logical Toffoli gates through their Clifford + $T$ decomposition.
Logical Clifford gates are implemented through the CliNR scheme~\cite{CliNR_original} executed at the logical level.
Recall that CliNR works in three steps: resource state preparation (RSP), resource state verification (RSV) and resource state injection (RSI) represented in \cref{fig:CliNR_circuit}(a).
\added{The efficiency of the protocol} originates from
(i) replacing a Clifford circuit by the offline preparation of a resource state (RSP), which can be prepared using error detection instead of error correction,
(ii) preparing and verifying the CliNR resource state using an MEDM, merging RSP and RSV,
(iii) implementing RSI through physical operations (transversal CNOT gates and destructive measurements)
to avoid the time penalty of RSI in the original CliNR scheme.

\subsection{CliNR Protocol for Physical Clifford Circuits}\label{sec:physical_CliNR}
The CliNR protocol applies a physical Clifford operation $\mathcal{C}$ by preparing a resource stabilizer state offline on auxiliary blocks A and B.
The auxiliary blocks are initialized in Bell states and $\mathcal{C}$ is applied to auxiliary block B using noisy gate-based operations. 
To validate preparation of the resource state, a selection of stabilizers is measured using an auxiliary qubit ($M_C$ in \Cref{fig:physical_CliNR}).
If any of the stabilizer measurements are non-zero, the protocol restarts.
Otherwise, the Clifford gate is applied via state injection using a transversal CNOT between the data block and auxiliary block A and measurements in the X and Z basis on the data block and auxiliary block A respectively. 
The output \added{$\ket{\mathcal{C}\psi}$} is teleported to auxiliary block B.

\subsection{Logical CliNR Protocol}\label{sec:logical_CliNR}
\added{We now describe the logical CliNR protocol in more detail for a logical Clifford operator with support on $\ell$ qubits acting on a code block with $k \ge \ell$ logical qubits. 
For simplicity, we present results in this section in terms of logical stabilizers denoted $P$ and denote the physical representative of a logical $P$ operator $\overline{P}$.
The resource state we prepare for logical CliNR is equivalent to preparing auxiliary blocks A and B in logical Bell pairs, then applying a logical $\mathcal{C}$ operation to block B. 

For the initial logical Bell pair state, the operators $X_{A,i}X_{B,i}$ and $Z_{A,i}Z_{B,i}$ for $0 \le i < \ell$ are logical stabilizers and the operators $Z_{A,i}$ and $X_{B,i}$ are logical destabilizers (i.e. they anticommute with exactly one of the logical stabilizers and commute with all others - see \cite{aaronson_improved_2004}).
The logical stabilizers after applying $\mathcal{C}$ to block B are found by conjugating by $\mathcal{C}_B$ and we denote these as $M_{X,i} ={\mathcal{C}}_B{X}_{A,i}{X}_{B,i}{\mathcal{C}}_B^{-1}$ and $M_{Z,i} ={\mathcal{C}}_B{Z}_{A,i}{Z}_{B,i}{\mathcal{C}}_B^{-1}$ for $0\le i<\ell$.

We measure physical representatives of each of the $2\ell$ commuting logical stabilizers  using cat-based measurements and employ either the multi-Pauli techniques of \Cref{sec:MEDM,sec:MECM} or the single-Pauli techniques of \Cref{sec:meas_intro} (block $\overline{M}_C$ in \Cref{fig:logical_CliNR}).
As these measurements are between two adjoining code blocks, stitched cat states may be required (see \cite{walking_cat}).
Let $\mathbf{x}_i$ and $\mathbf{z}_i$ be the measurement outcomes of the logical measurements ${M_{X,i}}$ and ${M_{Z,i}}$ for $0\le i < \ell$.
We correct into the $+1$ eigenspace of the logical stabilizers by applying a logical Pauli correction given by $R=\mathcal{C}_B\left(\prod_{0\le i<\ell}Z_{A,i}^{\mathbf{x}_i}X_{B,i}^{\mathbf{z}_i}\right)\mathcal{C}_B^{-1}$ which is a product of the logical destabilizers of the desired state (indicated by the block labeled $\overline{R}$ in \Cref{fig:logical_CliNR}).}

The logical Clifford operation is applied via state injection as follows. 
First, we apply physical CNOTs transversally between corresponding qubits in the data block and auxiliary block A.
For this step we assume that the two blocks are adjacent. 
For data blocks encoded in the same CSS code, this results in a logical CNOT operation between the two blocks.
Next, we perform destructive measurements on the data block in the $X$ basis.
As the data is encoded in a CSS code, this allows us to determine the value $\mathbf{x}_i$ of each logical $X_i$.
Similarly, we measure the auxiliary block A in the $Z$ basis giving a logical outcomes $\mathbf{z}_i$.
We then apply the logical Pauli operator $Q = \mathcal{C}\left(\prod_{0\le i<\ell}{X}_i^{\mathbf{z}_i}{Z}_i^{\mathbf{x}_i}\right)\mathcal{C}^{-1}$ to auxiliary block B ($\overline{Q}$ in \Cref{fig:CliNR_circuit}).
As a result, the output state $\ket{\mathcal{C}\psi}_L$ is teleported to auxiliary block B. 
In \Cref{sec:logical_CliNR_proof} we prove that the above protocol applies a logical $\mathcal{C}$ and show how the protocol works in combination with frame tracking.

\added{We now discuss the case where the logical Clifford to be applied has support on fewer qubits than the number of logical qubits in the code block.
In this case $\ell < k$ and a logical operation is only applied to $\ell$ of the $k$ logical qubits  but all $k$ logical qubits of the data block are teleported to the auxiliary block B. }

Logical CliNR  can be adapted to handle constraints on the maximum weight of logical Pauli measurements in the walking cat architecture (see \Cref{sec:clinr_accessibility}) and there is also a logical CZNR protocol for circuits composed of CZ and S gates which uses only one auxiliary code block (see \Cref{sec:logical_CZNR}).

\section{Numerical results}
In this section, we present simulation results for the MEDM and MECM protocols of \Cref{sec:MEDM,sec:MECM}, as well as the  logical CliNR protocol of \Cref{sec:logical_CliNR}.
In \cite{walking_cat} the authors demonstrate numerically that measurement flip errors for the Q70 and Q102 codes can be modeled as an independent bit flip channel with rate $p_F = C_1 \overline{w} p$ where $C_1= 2.1$ is a constant, $\overline{w}$ is the cat state size and $p$ is the error rate of 2-qubit gates for the device. 
For our results, we use a flip rate of $p_F = 6.3\times 10^{-3}$ corresponding to a physical error rate of $p=10^{-4}$ and cat state size $\overline{w}=30$.
We use the target logical error rate of $\varepsilon=10^{-10}$.
The metric we use for comparison is the number of cat-based measurements required by the protocol.

\begin{table}[]
    \centering
    \begin{tabular}{|l|r|}
    \hline
    Scenario & Saving\\
    \hline
    MEDM $\ell=10$  &   2.31x\\
    MECM $\ell=10$   & 2.41x\\
    \hline
    MEDM $\ell=20$   &  2.81x\\
    MECM $\ell=20$    & 2.96x\\
    \hline
    CliNR Toffoli Q70    &  4.2x\\
    CliNR Toffoli Q102    & 5.1x\\
    \hline
    CliNR Random Q70    & 18.6x\\
    CliNR Random Q102    & 73.1x\\
    \hline
    \end{tabular}
    \caption{Multi-Pauli MEDM and MECM protocols compared to single-Pauli Viterbi protocol for $\ell=10,20$ and logical CliNR protocol versus Viterbi gate-based  circuit synthesis. 
    For more detail, see \Cref{tab:MEDM,tab:MECM,tab:logical_CliNR}}
    \label{tab:medm_mecm_high_level}
\end{table}
In \Cref{tab:medm_mecm_high_level} we give a high-level summary for the MEDM, MECM, and logical CliNR protocols.
To achieve a logical error rate of $\varepsilon=10^{-10}$, the Viterbi protocol of \Cref{sec:meas_intro} requires almost 2.4 times as many cat-based measurements as the MECM protocol for measuring $\ell=10$ commuting Paulis and almost 3 times as many for $\ell=20$ commuting Paulis.
The MEDM protocol requires only 1.81 cat-based measurements per target Pauli and the MECM only 1.71 cat-based measurements (see \Cref{tab:MEDM,tab:MECM}).
For full details of simulation methodology, see \Cref{sec:MEDM_results}.

We then simulated logical CliNR using truncated MEDM for stabilizer state preparation.
We compared the number of cat-based measurements required for logical CliNR to the number required by the \textit{Viterbi gate-based  protocol}.
The Viterbi gate-based  protocol involves synthesizing a logical Clifford into logical single and two-qubit gates, which are then implemented via logical measurements. 
Each logical measurement is performed using the Viterbi measurement protocol (see \Cref{sec:cliffords_walking_cat}).
We focus on this gate-based compilation strategy for logical Clifford gates because it is the most widely adopted, to the best of our knowledge. 
Another strategy was recently proposed in~\cite{kliuchnikov2026clifford}.

We considered two scenarios - the Toffoli circuit requiring four logical qubits as set out in \Cref{fig:toffoli_circuit} and a \added{set of 100 randomly generated Clifford operators}. 
For each scenario, we simulated using the Q70 code with 6 logical qubits and the Q102 code with 22 logical qubits from \cite{walking_cat}.
We found that the Viterbi gate-based  protocol requires up to 5.1x more cat-based measurements than MEDM CliNR for Toffoli circuits.
For random Clifford circuits the advantage is even larger, and this scenario illustrates the scope of potential gains from the MEDM CliNR protocol.
For full details of simulation methodology and results see \Cref{sec:clinr_simulation}.

\added{In this analysis we focused on establishing the reliability of the classical logical-measurement decoding problem rather than including all possible sources of faults, for instance in the transversal CNOT, destructive measurements, data-block creation, and qubit transport.
The reason for this is that analysis of these effects is very device-dependent and the CliNR protocol is applicable to any device with access to transversal CNOT and cat-based measurements.
We also note the flexibility of the protocol and that accuracy can be tuned by modifying the target logical error rate if necessary.}

\section{Conclusion}\label{sec:conclusion}

We have presented methods to \added{improve the efficiency} of logical measurements, Clifford gates, and non-Clifford Toffoli gates in quantum LDPC codes.
In future work, it would be interesting to investigate which features of quantum LDPC codes are sufficient to ensure independence of cat-based measurement outcomes.
Another open question is the design of explicit families of scheduler codes instead of selecting codes with the best-known-distance.
Finally, the minimal space and time footprint of our logical operations makes them an ideal target for near-term experimental demonstration.

\section{Acknowledgments}

The authors would like to thank John Gamble for his support during the preparation of this manuscript, as well as  Edwin Tham, Felix Tripier and Aharon Brodutch for insightful discussions.

\appendix

\section{Logical Clifford Operators in the Walking Cat Architecture}\label{sec:cliffords_walking_cat}

\subsection{Clifford Operations via Frame Tracking}
In the walking cat architecture \cite{walking_cat}, logical Clifford operators are implemented where possible via \textit{frame tracking}. 
This method does not require any physical operations but instead involves maintaining a Pauli frame. 
For a code block with $k$ logical qubits, the Pauli frame is a tableau comprising a binary $2k \times 2k$ symplectic matrix and a length $2k$ vector which represents a  logical Pauli basis with signs (see \cite{aaronson_improved_2004}).
When a logical Clifford operator $\mathcal{C}$ is applied, the tableau is updated by conjugating the logical Pauli basis by $\mathcal{C}$.

Not all logical Clifford operations can be implemented via frame tracking and this means that they need to be implemented physically. 
For instance, frame tracking can only be applied efficiently for operations within the same code block.
Frame tracking also requires that each element of the logical Pauli basis is accessible in the sense that cat-based measurements are possible within architecture constraints. 

\subsection{Accessibility of Logical Pauli Measurements}
Here we describe the accessibility of logical Pauli measurements in more detail.
A key parameter of the walking cat architecture is the \textit{maximum cat state size} $\overline{w}$ which can be produced by the cat factory.
Within a code block, only Pauli operators of weight $\overline{w}$ or less can be measured using cat states. 
The \textit{minimum weight representative} of a logical Pauli with binary symplectic representation $\mathbf{v}$ for a quantum stabilizer code with check matrix $H$ is the element of the coset $\mathbf{v} +\braket{H}$ with minimum symplectic weight.
A logical Pauli measurement is \textit{accessible} if the Pauli has a minimum weight representative with weight $\overline{w}$ or less.
Accessibility depends on the QECC used - for Q70 all logical Paulis have representatives with weight $\le 18$ but for Q102 some logical Paulis have minimum representatives of weight 30.

Where all logical Cliffords are not guaranteed to be accessible, only single qubit logical Cliffords (generated by $\braket{H,S}$) and logical qubit permutations can be implemented via frame tracking.
In this case, we choose a low-weight logical basis $\mathcal{B}$ and work with the \textit{logical width} $w$ such that any product of $w$ or fewer elements of $\mathcal{B}$ is accessible.

\textit{CSS logical operators} are those composed of only X-type or Z-type logical operators. 
Enumerating these and checking for minimum weight representatives is far less challenging than for general logical Paulis because there are $2^{k+1}$ CSS logical operators versus $2^{2k}$ logical Paulis in total.
CSS logical operators usually have lower weight minimum representatives than non-CSS logical operators - for Q70 all CSS logical operators have representatives of weight 11 or less, and for Q102 the maximum CSS logical weight is 20.

\subsection{Viterbi Gate-Based  Protocol for Logical Clifford Operations}\label{sec:GB+viterbi}
\begin{figure}[hbt]
    \centering
    \includegraphics[width=\linewidth]{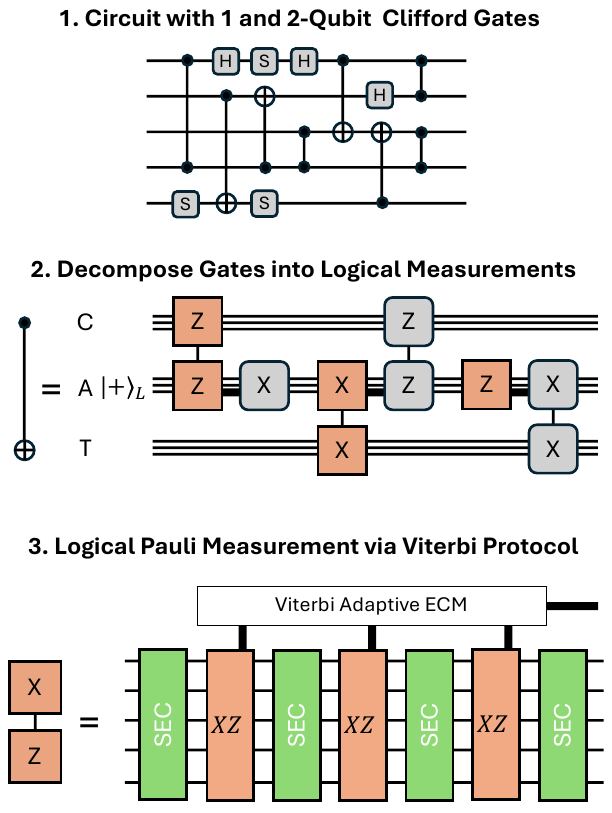}
    \caption{Viterbi Gate-Based  Protocol for implementing Clifford operations. We illustrate the protocol using a circuit composed of CNOT and CZ gates. Each logical CNOT between control qubit C and target qubit T is implemented using  a logical auxiliary qubit A and three Pauli measurements. Each Pauli measurement employs the Viterbi measurement protocol which requires on average 5.06 cat-based measurements for our default error rate parameters. }
    \label{fig:GB+viterbi}
\end{figure}

Where a Clifford operation cannot be implemented via frame tracking, the \textit{Viterbi gate-based  protocol} is used (see \Cref{fig:GB+viterbi}).
The input to this protocol is a logical Clifford operator, which we decompose into a circuit of one and two-qubit logical Clifford gates (see \cite{aaronson_improved_2004,synthesis_templates,random_clifford,webster2025heuristicoptimalsynthesiscnot}). 
Single-qubit logical Clifford gates can be implemented using two logical Pauli measurements, and two-qubit gates using three logical Pauli measurements as set out in \Cref{fig:GB+viterbi}.

Each logical Pauli measurement is implemented using cat-based measurements interleaved with SEC rounds using the Viterbi measurement protocol of \Cref{sec:meas_intro}.

\section{Classical Coding Techniques}
\subsection{Weight Enumerators, UER of Classical Codes and Posterior Probability Calculation}\label{sec:WE+UER}
In this section we explain how to calculate weight enumerators for cosets of binary linear codes. 
These are used to calculate the undetectable error rate and the posterior probability used in the MECM protocol of \Cref{sec:MECM}.

Given a binary vector $\mathbf{v}$ of length $m$ and full-rank  $\ell \times m$ binary matrix $G$, the \textit{weight enumerator} of the \textit{coset} $\mathbf{v}+\braket{G} = \{\mathbf{v} + \mathbf{u}G : \mathbf{u} \in \mathbb{F}_2^\ell\}$ is:
\begin{align}
    W_{\mathbf{v}+\braket{G}}(x,z) &= \sum_{\mathbf{e}\in \mathbf{v}+\braket{G}}x^{\wt{e}}z^{m-\wt{e}}\\
    &=\sum_{0\le i\le m}A_ix^iz^{m-i}.
\end{align}
The \textit{multiplicity} $A_i$ is the number of vectors of weight $i$ in the coset.
The weight enumerator can be calculated using $2^{\ell}$ binary vector operations using  Gray codes \cite{Gray_codes}.

The \textit{undetectable error rate} of the code $\braket{G}$ is the probability of the non-zero codewords and can be calculated by setting $\mathbf{v=0}$ and substituting in the flip/no flip probabilities for $x$ and $z$ as follows:
\begin{align}
     \UER{G}= W_{\braket{G}}(p_F,1-p_F) - P(\mathbf{0}).
\end{align}
We now show how to exactly calculate the posterior probability used as the termination condition for the MECM protocol of \Cref{sec:MECM}.
The probability of observing the outcome vector $\mathbf{v}$ is the sum of the probabilities of all errors which can produce the outcome $\mathbf{v}$.
This is equivalent to summing the probabilities of elements of the coset $\mathbf{v}+\braket{G}$ and so is given by: 
\begin{align}
    P(\mathbf{v}+\braket{G}) = W_{\mathbf{v}+\braket{G}}(p_F,1-p_F).
\end{align}
The \textit{posterior probability} of the logical outcome $\hat{\mathbf{u}}$ where $\hat{\mathbf{e}} = \mathbf{v} + \hat{\mathbf{u}}G$ is:
\begin{align}
    P(\hat{\mathbf{u}}|\mathbf{v}) = P(\hat{\mathbf{e}})/P(\mathbf{v}+\braket{G}).\label{eq:posterior_prob}
\end{align}

\subsection{Lookup Decoder and LER of Classical Codes}\label{sec:decoder+LER}
In this section, we show how to construct a lookup decoder for a binary linear code with $\ell \times m$ generator matrix $G$.
This produces an \textit{ideal maximum likelihood decoder} which is used for simulations of the MECM protocol and to calculate the logical error rate of binary linear codes exactly.
The decoder works by specifying a minimal coset leader $\hat{\mathbf{e}}$ for each coset $\mathbf{v}+\braket{G}$.
The lookup decoder applies the correction $\hat{\mathbf{e}}$ for any element of the coset.

The \textit{coset leader} is the element of the coset which is of minimum weight (and so highest probability) and minimum lexicographical order (so that we have a total order for each coset). 
There are $2^{m-\ell}$ coset leaders and these can be calculated in $\mathcal{O}(2^m)$ binary vector operations as follows.
Assuming the first $\ell$ columns of $G$ form an information set, we put $G$ into \textit{systematic form} using Gaussian elimination so that $G_S = [I|A]$ and $\braket{G_S} = \braket{G}$.
The \textit{complementary space} is generated by $C=[0|I_{m-\ell}]$.
Form the $m \times (2m-\ell)$ matrix $M = \begin{bmatrix}
    G_S&0\\C&I_{m-\ell}
\end{bmatrix}$ and iterate through all $2^m$ elements of $\braket{M}$. 
For each vector in the span, the coset is identified by the last $m-\ell$ columns and this allows us to track the best coset leader candidates.

The lookup decoder takes as input a binary vector $\mathbf{v}$ of length $m$. 
We identify the error coset by adding rows of $G_S$ to eliminate the first $\ell$ entries of $\mathbf{v}$ and this allows us to identify the coset by the last $m-\ell$ entries of the resulting vector.
The correction is the coset leader of the relevant coset.

This process allows us to identify the \textit{correctable errors} for the code $\braket{G}$ using an ideal decoder.
The decoder successfully corrects any error which corresponds to a coset leader, but any other error gives a logical error. 
Hence, we  choose the $2^{m-\ell}$ coset leaders as the set of correctable errors $CE_{\braket{G}}$ and the logical error rate is $\LER{G} = 1-P(CE_{\braket{G}})$.

\subsection{Heuristic Enumeration of Cosets Using Random Information Sets}\label{sec:WE_heuristic}
The exact methods for calculating coset probabilities of \Cref{sec:WE+UER} have exponential complexity in $\ell, m$.
For large codes, we instead use heuristic methods. 
The random information set algorithm for distance-finding \cite{Leon_1988,Coffey_Goodman_1990,Dumer_Kovalev_Pryadko_2016} can be modified to sample from low-weight elements of the coset $\mathbf{v}+\braket{G}$.
As low-weight elements have the highest contribution to the coset probability, this gives an estimate of the coset probability.
By returning the lowest weight element with least lexicographic order, this method can also be used as a heuristic decoder. 
The method is set out in \Cref{alg:heuristic_coset}.
The function $\texttt{RREF}(G_v,\texttt{ix})$ performs reduced row echelon using the column order specified by the vector $\texttt{ix}$ and on average returns low weight vectors (see \cite{webster2026distancefindingalgorithmsquantumcodes}). 
The $\texttt{wLexMin}$ function returns the minimal vector by weight then lexicographic order.

\begin{algorithm}[hbt]
\caption{Heuristic Coset Enumerator}\label{alg:heuristic_coset}
\KwData{$\ell \times m$ generator matrix $G$; Length $m$ vector $\mathbf{v}$; Number of iterations $N >1$}
\KwResult{Set of low weight vectors $V$; Coset leader $\hat{\mathbf{e}}$}
$V := \texttt{set}()$\;
$\hat{\mathbf{e}} := \texttt{None}$\;
$G_v := \begin{bmatrix}\mathbf{v}&1\\G&0\end{bmatrix}$\;
\For{$i \in [1..N]$}{
    $\texttt{ix} := \texttt{RandomPermutation}(m)$\;
    $U := \texttt{RREF}(G_v,\texttt{ix})$\;
    \For{$\mathbf{u} \in U$}{
        \If{$\mathbf{u}[m] = 1$}{
            $\mathbf{u} := \mathbf{u}[:m]$\;
            $\hat{\mathbf{e}} := \texttt{wLexMin}(\hat{\mathbf{e}},\mathbf{u})$\;
            $V.\texttt{add}(\mathbf{u})$\;
        }
    }
}
\end{algorithm}

\section{Measurement of Commuting Paulis}
\subsection{Modeling the MEDM and MECM Protocols}\label{sec:MEDM_results}
In this section we describe the methodology used to model the MEDM and MECM protocols of \Cref{sec:MEDM,sec:MECM}.
The first step of modeling MECM is to choose a scheduler code (see \Cref{sec:scheduler_codes}).
For $\ell \in [2,4,..20]$, we pre-calculated the weight enumerators of the best-known-distance  binary linear codes from \cite{codetables} with block size up to $m=64$ and $\ell$ logical bits. 
Substituting the flip/no flip probability using the method in \Cref{sec:WE+UER}, we calculated the undetectable error rate and selected the code with lowest $m$ which met the UER requirement of \Cref{eq:UER_requirement}.
We then used the closed form expressions for $m_N$ and $\subtext{N}{Avg}$ as set out in \Cref{sec:MEDM} to produce the results in \Cref{tab:MEDM}.

\begin{table}[hbt]
    \centering
    \begin{tabular}{|r|	r|	r|	r|	r|	r|	r|}
    \hline
$\ell$&	$m$&	$m_N$&	$\subtext{N}{Avg}$&	 $\subtext{m}{Avg}$&	$\subtext{m}{Avg}/\ell$&	{\makecell[r]{Viterbi/\\MEDM}}\\
\hline
1&	5&	4.9&	1.03&	5.1&	5.10&	0.99x\\
2&	8&	7.9&	1.05&	8.3&	4.13&	1.22x\\
4&	11&	10.7&	1.07&	11.5&	2.88&	1.76x\\
6&	15&	14.5&	1.10&	16.0&	2.66&	1.90x\\
8&	17&	16.4&	1.11&	18.3&	2.29&	2.21x\\
10&	20&	19.3&	1.13&	21.9&	2.19&	2.32x\\
12&	22&	21.1&	1.15&	24.3&	2.02&	2.50x\\
14&	24&	23.1&	1.16&	26.9&	1.92&	2.64x\\
16&	27&	25.9&	1.19&	30.7&	1.92&	2.64x\\
18&	29&	27.8&	1.20&	33.4&	1.85&	2.73x\\
20&	31&	29.7&	1.22&	36.1&	1.81&	2.80x\\
\hline
    \end{tabular}
    \caption{Results for Truncated MEDM Protocol: In this table, we give results for the MEDM protocol for $\ell \in [1,2,4,..20]$.
    The data in the table are as follows - $m=|G|$:  number of columns of $G$; $m_N$: average number of cat-based measurements per attempt; $\subtext{N}{Avg}$:  average number of attempts; MEDM $\subtext{m}{Avg} = m_N \subtext{N}{Avg}$:  average number of cat-based measurements for truncated MEDM;  $\subtext{m}{Avg}/\ell$ the average number of cat-based measurements per target Pauli; Viterbi/MEDM the number of cat-based measurements for the Viterbi protocol for a single Pauli (5.06) divided by the number required for MEDM. }
    \label{tab:MEDM}
\end{table}

\begin{table}[hbt]
    \centering
    \begin{tabular}{|r|	r|	r|	r|	r|	r|	r|r|}
    \hline
$\ell$&	$|G_0|$&	$|G_1|$& $m$	&$d$& $\subtext{m}{Avg}$&	$\subtext{m}{Avg}/\ell$&	{\makecell[r]{Viterbi/\\MECM}}\\
\hline
1&	5&	6&	11&	11&	5.1&	5.06x&	1.00x\\
2&	8&	12&	20&	13&	8.2&	4.11x&	1.23x\\
4&	11&	20&	31&	15&	11.3&	2.83x&	1.79x\\
6&	15&	22&	37&	15&	15.5&	2.58x&	1.96x\\
8&	17&	25&	42&	15&	17.8&	2.22x&	2.28x\\
10&	20&	30&	50&	17&	21.0&	2.10x&	2.41x\\
12&	22&	31&	53&	16&	23.4&	1.95x&	2.60x\\
14&	24&	36&	60&	17&	25.7&	1.84x&	2.76x\\
16&	27&	39&	66&	17&	29.1&	1.82x&	2.78x\\
18&	29&	41&	70&	17&	31.6&	1.76x&	2.88x\\
20&	31&	51&	82&	19&	34.2&	1.71x&	2.96x\\
\hline
    \end{tabular}
    \caption{Results for Truncated MECM Protocol: In this table, we give results for the MECM protocol for $1 \le \ell \le 20$.
    The data in the table are as follows: $|G_0|, |G_1|$ and $m=|G_0|+ |G_1|$ the number of columns of $G_0,G_1$ and $G$ respectively; $d$ the distance of the code $\braket{G}$; $\subtext{m}{Avg}$ the average number of cat-based measurements for truncated MECM;  $\subtext{m}{Avg}/\ell$ the average number of cat-based measurements per target Pauli; Viterbi/MECM the number of cat-based measurements for the Viterbi protocol for a single Pauli (5.06) divided by the number required for MECM. Note that $m_0=|G_0|$ gives a good approximation for $\subtext{m}{Avg}$ (see \Cref{sec:scheduler_choice}).}
    \label{tab:MECM}
\end{table}

Modeling the MECM protocol of \Cref{sec:MECM} involved the following steps for $\ell \in [2,4,..20]$. 
We first chose the same code used for the MEDM protocol for $G_0$ (see \Cref{sec:scheduler_choice}).
We then used \Cref{eq:approx_LER} to approximate the LER for the codes with $\ell$ logical bits and this gave us an estimate for the distance $d$ required for $G = [G_0|G_1]$ to meet the LER requirement.
To satisfy the LER requirement, we chose the smallest code from \cite{codetables} with $\ell$ logical bits with distance $d-d_0$ where $d_0$ is the distance of $\braket{G_0}$.

We  modeled the performance of the MECM protocol using a variation of the lookup decoder of \Cref{sec:decoder+LER}.
As the block size $m$ of $G$ was in general too large for calculating the lookup decoder exactly, we sampled error strings $\mathbf{e}$ of weight between 0 and 7.
We decoded each $\mathbf{e}$ by enumerating all elements of the coset $\mathbf{e}+\braket{G}$ - as  $\ell\le 20$ for our data set, this could be done within a reasonable time frame. 
For $2 \le \ell \le 12$, we enumerated all errors of weight less than 7 to estimate $\subtext{m}{Avg}$ and to check the logical error rate (by summing the probability of errors leading to logical errors). Errors of weight greater than 7 make only a small contribution to these. 
For $\ell \ge 14$, we sampled a maximum of $4096$ errors of each weight to estimate $\subtext{m}{Avg}$, but did not verify the LER explicitly.
The results of the MECM simulation are set out in \Cref{tab:MECM}.

In the walking cat architecture, we would interleave each measurement with an SEC round and may require a final SEC at the end of the protocol, but we do not explicitly consider these here. Measurements may also be subject to a cat missing error but for simplicity we omit this effect in this work. This has only a small effect on the number of cat-based measurements required because cat state preparation rarely fails~\cite{walking_cat}.

Both protocols have relative advantages and are likely to be useful in different contexts.
The MECM protocol can be used directly on logical blocks without restarting and has a slightly lower average number of cat-based  measurements. 
On the other hand, the MEDM protocol does not require a decoder so it is less complex to model and run. 
Selecting an optimal generator matrix for the MEDM protocol depends on the undetectable error rate which is less complex to calculate exactly and we have closed form expressions for the key $\subtext{N}{Avg}, m_N$ and $\subtext{m}{Avg}$ values.

\subsection{MEDM/MECM Example - Repetition Code for Single Pauli Measurements}
In this example, we show how MEDM and MECM apply for single Pauli measurements ($\ell = 1$) and recover the results of \cite{walking_cat} for the EDM, ECM and Viterbi measurement protocols.
To apply the MEDM protocol, we require $\braket{G}$ to meet $U_{\braket{G}}=\UER{G}/(P(\mathbf{0})+\UER{G}) < \varepsilon=10^{-10}$.
The codewords of the length $m$ repetition code are the all zeros vector and the all ones vector so:
\begin{align*}
    \UER{G}&= P(\mathbf{1}) = p_F^m;\\
    P(\mathbf{0})&=(1-p_F)^m.
\end{align*}

\added{
\begin{table}[]
    \centering
    \begin{tabular}{|l|r|r|r|r|r|}
\hline
    m & 2& 3 & 4 & 5 & 6 \\
    \hline
    $U_{\braket{G}}$ & $\expnumber{4.0}{-5}$& $\expnumber{2.5}{-7}$  &$\expnumber{1.6}{-9} $  & $\expnumber{1.0}{-11}$  & $\expnumber{6.3}{-14} $\\
    \hline
    \end{tabular}
    \caption{Undetectable Error Rate (UER) for repetition codes with block length $m\in [2,6]$ and $p_F=6.3\times10^{-3}$.}
    \label{tab:UER_repetition_code}
\end{table}
}
In \Cref{tab:UER_repetition_code}, we see that the smallest $m$ meeting the UER requirement is $m=5$.
The average number of attempts required to meet the target logical error rate is found by applying \Cref{eq:Ravg}:
\begin{align*}
    \subtext{N}{Avg} = 1/(P(\mathbf{0})+\UER{G}) = 1/((1-p_F)^m+p_F^m)
\end{align*}
For $m=5$ this gives a value of $\subtext{N}{Avg}\approx 1.03$ so the average number of cat-based measurements for the full MEDM protocol is $5.16$ which closely matches the results in \cite{walking_cat} for the EDM protocol.
For the truncated MEDM protocol, we calculate the average number of cat-based measurements per round using \Cref{eq:mR} using $\ell=1$ for the length $m$ repetition code as follows:
\begin{align*}
    m_N &= 1 + 1 + \sum_{2 \le i < m}(P(\mathbf{0})+\UER{G|i})\\
    &= 2 + \sum_{2 \le i < m}((1-p_F)^i+p_F^i).
\end{align*}
For $m=5$, this gives a value of $m_N\approx 4.9$ so that the average number of cat-based measurements for the truncated MEDM protocol is $\subtext{m}{Avg} = m_N \times \subtext{N}{Avg} \approx 5.1$.

We now consider the MECM protocol using a length $m$ repetition code and a majority vote decoder.
If $m$ is odd, the correctable errors are those with weight $(m-1)/2$ or less.
Hence, the logical error rate is:
\begin{align*}
    \LER{G} =& 1-\sum_{0 \le i \le (m-1)/2}\binom{m}{i} p_F^i(1-p_F)^{m-i}.
\end{align*}
The logical error rate requirement requires a larger repetition code to meet the target - in \Cref{tab:LER_repetition_code} we see that the smallest code meeting the LER target is $m=11$ and this matches the value found for the EDM protocol in \cite{walking_cat}. 
\added{
\begin{table}[]
    \centering
    \begin{tabular}{|l|r|r|r|r|r|}
\hline
    m & 9  & 11  & 13\\
    \hline
    LER & $\expnumber{1.2}{-9}$ &   $\expnumber{2.8}{-11}$ & $\expnumber{6.5}{-13}$\\
    \hline
    \end{tabular}
    \caption{Logical Error Rate (LER) for repetition codes with block length $m\in [9..13]$ and $p_F=6.3\times 10^{-3}$.}
    \label{tab:LER_repetition_code}
\end{table}
}
For the truncated MECM protocol, assume that the measurement outcome vector $\mathbf{v}$ has $m_1=\wt{v} > m/2$ ones so that the most likely outcome $\hat{\mathbf{u}}=1$ with correction $\hat{\mathbf{e}} = \mathbf{1+v}$.
The termination condition of \Cref{eq:posterior_prob} can be written 
\begin{align*}
P(\hat{\mathbf{u}}|\mathbf{v}) &= P(\hat{\mathbf{e}})/P(\mathbf{v}+\braket{G})\\
&= P(\mathbf{v+1})/(P(\mathbf{v}) + P(\mathbf{v+1}))\\
&= [1+P(\mathbf{v})/P(\mathbf{v+1})]^{-1}\\
&= [1 + p_F^{m_1}(1-p_F)^{m-m_1}p_F^{m_1-m}(1-p_F)^{-m_1}]^{-1}\\
&= [1 + (p_F/(1-p_F))^{2m_1-m}]^{-1}\\
&= [1 + (p_F^{-1}-1)^{m-2m_1}]^{-1}.\\
\end{align*}
The posterior probability is less than the target logical error rate when:
\begin{align*}
\varepsilon^{-1} -1 < (p_F^{-1}-1)^{m-2m_1}.
\end{align*}
or equivalently:
\begin{align*}
{2m_1-m} > \log(\varepsilon^{-1} -1)/\log(p_F^{-1}-1).
\end{align*}
The case where $m_1 < m/2$ leads to a similar condition on $m-2m_1$, so we obtain the termination condition:
\begin{align*}
|2m_1-m| > \log(\varepsilon^{-1} -1)/\log(p_F^{-1}-1).
\end{align*}
\subsection{MEDM/MECM Example - Four Pauli Measurements}
We now work through the MEDM and MECM protocols for $\ell=4$ Pauli measurements. 
For the MEDM protocol, we calculate the undetectable error rate and the condition of \Cref{eq:UER_requirement} of the best-known-distance binary linear codes from \cite{codetables} for $\ell=4$ and find that the  block length 11 code is the smallest code meeting the UER requirement of \Cref{eq:UER_requirement} and has generator matrix:
\begin{align*}
    G = \begin{bmatrix}
10001011101\\
01001101011\\
00101110111\\
00010001111
    \end{bmatrix}.
\end{align*}
\added{
\begin{table}[]
    \centering
    \begin{tabular}{|l|r|r|r|r|r|}
    \hline
    m & 7 & 8 & 10 & 11 & 12 \\
    \hline
    $U_{\braket{G}}$ & $\expnumber{1.8}{-6}$& $\expnumber{2.3}{-8}$  &$\expnumber{1.6}{-8} $  & $\expnumber{6.2}{-11}$  & $\expnumber{7.8}{-13}$ \\
    \hline
    \end{tabular}
    \caption{Undetectable Error Rate (UER) for best-known-distance codes with $\ell=4$ logical bits and block length $m\in [7..12]$ and $p_F=6.3\times 10^{-3}$.}
    \label{tab:UER_l_4}
\end{table}
}
Each row of $G$ corresponds to one of the Paulis $P_0, P_1, P_2$ and $P_3$ we wish to measure.
Each column of $G$ corresponds to a product of these Paulis.
The first 4 columns have weight 1 and so correspond to products of a single Pauli. 
The fifth column corresponds to the product $P_0P_1P_2$ and the sixth column $P_1P_2$.

To find the average number of attempts for the MEDM protocol using $G$, we use \Cref{eq:Ravg} and the UER of the code and find:
\begin{align}
    \subtext{N}{Avg} = (\UER{G}+ (1-p_F)^{11})^{-1} \approx 1.07
\end{align}
For the truncated MEDM protocol, we calculate the average number of cat-based measurements per attempt by calculating the UERs of the codes specified by the first $i=[5,..,10]$ columns of $G$ and using \Cref{eq:mR} as follows:
\begin{align*}
    m_N &= 5 + \sum_{5\le i\le 10}(\UER{G|i} + (1-p_F)^i) \approx  10.75.
\end{align*}
Note that for $\ell=4$, finding the UER involves calculating the weight enumerator as set out in \Cref{sec:WE+UER} and involves $2^\ell = 16$ binary vector additions.
For the truncated MEDM protocol, the average number of cat-based measurements is $\subtext{m}{Avg} = \subtext{N}{Avg}m_N \approx 11.5$.

For the MECM protocol, in \Cref{tab:LER_l_4} we estimate the logical error rate of the best-known-distance binary linear codes from \cite{codetables} for $\ell=4$ using \Cref{eq:approx_LER}. 
We find that the block length 29 code with distance 15 is the smallest code with an LER upper bound estimate  meeting the requirement. 

\begin{table}[]
    \centering
    \begin{tabular}{|l|r|r|r|r|r|}
    \hline
    m & 26 & 27 & 28 & 29 & 30 \\
    \hline
    d & 13 & 14 & 14 & 15 & 16 \\
    \hline
    $\textrm{LER}$ & $\expnumber{2.3}{-10}$&$ \expnumber{3.1}{-10}$  &$\expnumber{4.2}{-10}$   & $\expnumber{9.5}{-12}$  & $\expnumber{1.3}{-11}$ \\
    \hline
    \end{tabular}
    \caption{Logical Error Rate (LER) and distance for best-known-distance codes with $\ell=4$ logical bits and block length $m\in [26..30]$ and $p_F=6.3\times 10^{-3}$.}
    \label{tab:LER_l_4}
\end{table}

This code might not give the best average number of cat-based measurements $\subtext{m}{Avg}$ for the truncated MECM protocol. 
To optimize $\subtext{m}{Avg}$, we follow the method in \Cref{sec:scheduler_choice} and use $G= [G_0|G_1]$ where $G_0$ is the block length $m_0=11$ code used for the MEDM protocol which has distance $d_0=5$. 
To achieve the required LER, we use $G$ with $d \ge 15$.
Accordingly, we select for $G_1$ the smallest $\ell=4$ code with distance $d_1=10$ which has block size $m_1=20$. 
The resulting code $G= [G_0|G_1]$ has distance $d=d_0+d_1 = 15$ and block size $m=m_0 + m_1 = 31$.

To verify the truncated MECM protocol, we simulate all errors $\mathbf{e}$ with weight $\wt{e} \le 7$ and check that the logical error rate of the protocol is at the target logical error rate $\varepsilon = 10^{-10}$ and the average number of cat-based measurements. 
This weight is chosen because the probability of errors with higher weight is small enough that they have very little impact on the LER and $\subtext{m}{Avg}$.
We find the minimum weight correction $\hat{\mathbf{e}} = \mathbf{v} + \hat{\mathbf{u}}G$ by adding all possible linear combinations of $G$, which requires $2^4=16$ binary vector additions. 
This yields a logical error rate for the protocol of $5.7\times 10^{-11}<\varepsilon$ and an average number of cat-based measurements $\subtext{m}{Avg}\approx 11.3$ which is close to the optimal $m_0=11$.

\subsection{Accommodating Pauli Measurement Accessibility in MEDM/MECM}\label{sec:MEDM_MECM_accessibility}
We now show how to adapt the MEDM and MECM protocols of \Cref{sec:MEDM,sec:MECM} to take into account accessibility constraints set out in \Cref{sec:cliffords_walking_cat} for logical Pauli measurements.
We assume that a logical Pauli basis $\mathcal{B}$ has been selected and that any product of $w$ basis elements is accessible.
Let $P_i, 0 \le i<\ell$ be a set of independent commuting Paulis we wish to measure. 
Any product of the $P_i$ can be expressed uniquely as a product of elements of $\mathcal{B}$ and stabilizer generators of the code. 
The logical weight is the number of elements of $\mathcal{B}$ in the product.

To ensure accessibility, we choose an $\ell \times m$ scheduler matrix $G$ whose columns represent accessible Paulis.
To maximize the efficiency of the MEDM and MECM schemes, the code generated by $G$ should have a low undetectable error rate.
We first enumerate all elements of the group generated by the target Paulis and calculate their logical weight. 
Products of the $P_i$ are represented as length $\ell$ binary vectors in the same way as set out in \Cref{sec:scheduler_codes}.
Let $C_w$ be the set of binary vectors representing products of the $P_i$ which have logical weight at most $w$.
By choosing the columns of $G$ from the vectors in $C_w$, we ensure that all measurements in the schedule are accessible.

The next step is to choose $\ell$ vectors from $C_w$ which form an information set.
This is done using matroid partitioning techniques \cite{edmonds_minimum_1965},\cite{terao_faster_2025} and if no such information set exists, the algorithm fails.
Otherwise, let $\mathbf{u}$ be the integer vector indexing the vectors in $C_w$ which form the information set.

The remaining $m-\ell$ columns of $G$ can now be chosen to give a low undetectable error rate.
We use an evolutionary algorithm based on the method in \cite{evol_alg_QECC}.
The individuals in the algorithm are represented as integer vectors of length $m-\ell$ which index the rows of $C_w$.
The initial population is a set of randomly generated vectors $\mathbf{v}_i$.
The fitness function for $\mathbf{v}_i$ is $1-\UER{G_i}$ where the columns of $G_i$ are  the columns of $C_w$ indexed by $\mathbf{u|v}_i$.
In each round, we select the $\mathbf{v}_i$ with highest fitness.
Mutation is done by choosing a entry in $\mathbf{v}_i$ at random and replacing it with an integer in $[0,..,|C_w|-1]$.
We also considered greedy and A* algorithms for optimization but found these challenging to apply due to the large number of columns typically required for the scheduler matrices.

\section{Logical Clifford Operations}

\subsection{Logical CliNR Proof}\label{sec:logical_CliNR_proof}
Here we prove that the logical CliNR protocol outlined in \Cref{sec:logical_CliNR} has the desired logical action and show how to adapt the protocol to accommodate frame tracking.

To verify the protocol, we check that each element of the stabilizer group $\mathcal{S}$ of the QECC stabilizes the output block B throughout the protocol.
We then verify the logical action by tracking how logical $X_i$ and logical $Z_i$ in the data block D is mapped to block B.

Firstly, we note that both auxiliary blocks A and B are prepared in a logical $\ket{0}$ states so they are stabilized by each element of $\mathcal{S}$ acting on their respective blocks. 
The measurements in $M_C$ are logical measurements and SEC rounds are done in between these which ensure that block B remains in the codespace.
The corrections $\overline{R}$ and $\overline{Q}$ are both products of logical Pauli operators and so do not take block B out of the codespace.
Hence, block B is stabilized by elements of $\mathcal{S}$ throughout the protocol.

Next, we assume that the data block is in the $+1$ eigenspace of logical $Z_{D,i}$ and consider logical stabilizers of the logical CliNR circuit after each step.
After state preparation, the logical stabilizers are given by $Z_{D,i}, X_{A,j}C_BX_{B,j}C_B^{-1}$ and $Z_{A,j}C_BZ_{B,j}C_B^{-1}$ for $0 \le j < \ell$.
The transversal CNOT maps $Z_{A,j}$ to $Z_{D,j}Z_{A,j}$ and so the resulting logical stabilizers are: $Z_{D,i}, X_{A,j}C_BX_{B,j}C_B^{-1}$ and $Z_{D,j}Z_{A,j}C_BZ_{B,j}C_B^{-1}$ for $0 \le j < \ell$.
We then measure all qubits on the data block in the $X$ basis. 
Providing the logical Paulis are CSS (i.e. strings of either physical $X$ or $Z$ operators) we can infer the value $\mathbf{x}_i$ of each logical $X_i$. 
The resulting logical stabilizers are: $(-1)^{\mathbf{x}_j}X_{D,j},X_{A,j}C_BX_{B,j}C_B^{-1}$ for $0 \le j <\ell$ and $Z_{A,i}C_BZ_{B,i}C_B^{-1}$.
After measuring auxiliary block A in the Z basis and inferring the values of the logical $Z$ operators in the length $\ell$ binary vector $\mathbf{z}$, the logical stabilizers are: $(-1)^{\mathbf{x}_j}X_{D,j},(-1)^{\mathbf{z}_j}Z_{A,j}$ for $0\le j <\ell$ and $(-1)^{\mathbf{z}_i}C_BZ_{B,i}C_B^{-1}$.
We then apply $Q = C\left(\prod_j X_j^{\mathbf{z}_j} Z_j^{\mathbf{x}_j}\right) C^{-1}$ to auxiliary block B.
This has the effect of canceling the sign on the stabilizer $(-1)^{\mathbf{z}_i}C_BZ_{B,i}C_B^{-1}$ and results in the following stabilizers $(-1)^{\mathbf{x}_j}X_{D,j},(-1)^{\mathbf{z}_j}Z_{A,j}$ for $0\le j <\ell$ and $C_BZ_{B,i}C_B^{-1}$.
Hence, the stabilizer $Z_{D,i}$ on the data block is mapped to $C_BZ_{B,i}C_B^{-1}$ on auxiliary block B.

A similar argument shows that $X_{D,i}$ is mapped to $C_BX_{B,i}C_B^{-1}$ and so the logical action for logical CliNR is as claimed.

We now consider the situation where logical CliNR is applied in the middle of an operation where logical Cliffords have been applied via frame tracking (see \Cref{sec:cliffords_walking_cat}).
In this case, let $\mathcal{F}$ be the accumulated Clifford operations applied via frame tracking.
Via frame tracking, the logical Pauli X operators have been mapped to $\mathcal{F}X_i\mathcal{F}^{-1}$ and $\mathcal{F}Z_i\mathcal{F}^{-1}$.
These are not guaranteed to be CSS logical Paulis so the measurement of physical X and Z operators may not suffice to infer the logical X and Z measurement values required by the protocol.
The solution is to absorb frame tracking into the Clifford $\mathcal{C}$ to be applied via logical CliNR so that we implement $\mathcal{CF}$. 
This ensures that we are working in the original CSS Pauli basis and means that logical X and Z measurement values can be inferred from physical X and Z measurements.
After applying logical CliNR in this way, the Pauli frame has been `cleared' and we are again working in the original CSS Pauli basis.

\subsection{Logical CZNR Protocol}\label{sec:logical_CZNR}
\begin{figure}[hbt]
    \centering
    \begin{subfigure}[t]{0.23\textwidth}
        \centering
        \includegraphics[width=\textwidth]{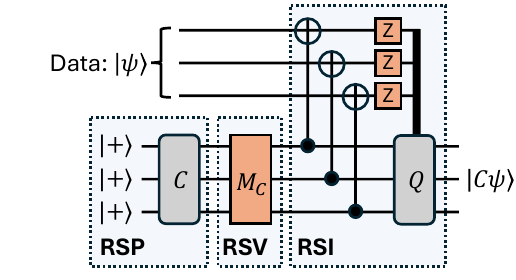}
        \caption{Physical CZNR}\label{fig:physical_CZNR}
    \end{subfigure}%
    ~
    \begin{subfigure}[t]{0.23\textwidth}
        \centering
        \includegraphics[width=\textwidth]{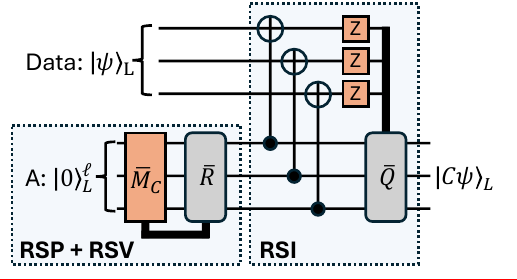}
        \caption{Logical CZNR}\label{fig:logical_CZNR}
    \end{subfigure}
    \caption{Physical and Logical CZNR Circuits: see main text for an explanation of the notation used here.}
    \label{fig:CZNR_circuit}
\end{figure}

The logical CZNR protocol can be used for Clifford circuits composed of CZ and S gates and is set out in \Cref{fig:CZNR_circuit}. 
Logical CZNR is simpler to execute than the logical CliNR protocol of \Cref{sec:logical_CliNR} - only one auxiliary code block is required and so only we execute only one transversal CNOT and one destructive Z measurement. 
The stabilizers of the injected state only need to be measured over one code block so stitched cat states are not required.

The circuits for the physical and logical CZNR protocols are set out in \Cref{fig:CZNR_circuit}.
The physical CZNR protocol of \Cref{fig:physical_CZNR} prepares a graph state in auxiliary block A by applying a diagonal Clifford operator $\mathcal{C}$ composed of S and CZ gates to physical $\ket{+}$ states (note that an S gate produces an edge from a vertex to itself).
As for CliNR, the graph state is validated by measuring stabilizers and the Clifford operation applied via state injection.

For logical CZNR we apply a logical diagonal Clifford $\mathcal{C}$ to an input logical state ${\ket{\psi}}_L$.
The logical resource state is the result of applying $\mathcal{C}$ to the auxiliary block prepared in logical $\ket{+}$.
The logical $\ket{+}$ state has logical stabilizers $X_i$ and destabilizers $Z_i$ for $0\le i<\ell$.
After applying logical $\mathcal{C}$, the logical stabilizers are $M_{i} ={\mathcal{C}}{X}_{i}{\mathcal{C}}^{-1}$ for $0\le i<\ell$.
The $M_{i}$ are measured using either the multi-Pauli techniques of \Cref{sec:MEDM,sec:MECM} or by measuring each stabilizer using the single-Pauli techniques of \Cref{sec:meas_intro} ($M_C$ in \Cref{fig:CZNR_circuit}).
Let $\mathbf{u}_i$ be the logical measurement outcome of $M_{i}$.
We correct into the $+1$ eigenspace of $M_i$ by applying the logical Pauli correction $R=\prod_{0\le i<\ell}Z_{i}^{\mathbf{u}_{i}}$ which is a product of destabilizers of the desired state  ($\overline{R}$ in \Cref{fig:CZNR_circuit} - note that as $\mathcal{C}$ is diagonal, $\mathcal{C}Z_i\mathcal{C}^{-1} = Z_i$).

We then apply a transversal CNOT between the  auxiliary block and the data block and perform destructive measurements on the data block in the Z basis.
This allows us to determine the values of each of the $\ell$ logical $Z$ operators in the data block, yielding a length $\ell$ logical outcome string $\mathbf{m}$.
We then apply the logical Pauli correction $Q = \mathcal{C}\left(\prod_{0\le i<\ell}{X}_i^{\mathbf{m}_i}\right)\mathcal{C}^{-1}$ to the auxiliary block ($\overline{Q}$ in \Cref{fig:CZNR_circuit}).
The output $\ket{\mathcal{C}\psi}_L$ is teleported to auxiliary block. 

\subsection{Accommodating Pauli Measurement Accessibility in Logical CliNR}\label{sec:clinr_accessibility}
We now show how to adapt the logical CliNR protocol of \Cref{sec:logical_CliNR} to take into account accessibility constraints set out in \Cref{sec:cliffords_walking_cat} for logical Pauli measurements.
We assume logical width $w$ so that any product of $w$ elements of the logical Pauli basis is accessible and that any CSS logical operator is accessible (see \Cref{sec:cliffords_walking_cat}).

The first step is to use  Aaronson and Gottesman Clifford circuit synthesis \cite{aaronson_improved_2004} to break the desired logical Clifford into three types of layers - those including only $S$, only Hadamard and only CNOT operators.
The maximum number of layers in such a decomposition is 11, but many Clifford circuits of interest can be implemented in far fewer layers.
We now show how to implement each type of layer in a way which guarantees that each stabilizer measurement in the CliNR protocol is accessible.

To implement a circuit involving only $S$ operators, we use the CZNR protocol of \Cref{sec:logical_CZNR}.
To prepare the stabilizer state on the auxiliary block, we must measure logical Paulis of form $S_iX_iS_i^{-1} =Y_i$. 
Measurements can be done one at a time using the Viterbi protocol or jointly using MEDM/MECM using a scheduler matrix with maximum column weight $w$.

To implement a circuit involving only $H$ operators, we use the CliNR protocol of \Cref{sec:logical_CliNR}. 
The stabilizers to be measured in step $M_C$ are of form $M_{X,i}=X_{A,i}Z_{B,i}$ and $M_{Z,i}=Z_{A,i}X_{B,i}$.
Any product of the $M_{X,i}$ is a CSS logical when restricted to auxiliary block A or block B.
CSS Paulis are accessible by assumption so these can be measured jointly using MEDM/MECM. 
The same applies to the $M_{Z,i}$ so we apply the MEDM/MECM protocol separately to products of the $M_{Z,i}$.

To implement a CNOT circuit using the CliNR protocol, we note that any CNOT circuit can be represented in symplectic form as:
\begin{align}
    \textrm{Sym}_C = \begin{bmatrix}
        U&0\\0&U^{-T}
    \end{bmatrix},
\end{align}
where $U$ is a $\ell\times \ell$  invertible binary matrix and $U^{-T}$ is the inverse transpose (see for example \cite{Murphy_Kissinger_2023} for how to calculate the parity matrix $U$).
Accordingly, the stabilizers to be measured in step $M_C$ are CSS logical operators because they are of form $M_{X,i}=X_{A,i}\prod_{0\le j <\ell} X_{B,j}^{U[i,j]}$ and $M_{Z,i}=Z_{A,i}\prod_{0\le j <\ell}Z_{B,j}^{U^{-T}[i,j]}$.
Hence, they are accessible by assumption.

\subsection{Simulation of Logical CliNR}\label{sec:clinr_simulation}

\begin{figure}[hbt]
    \centering
    \includegraphics[width=\linewidth]{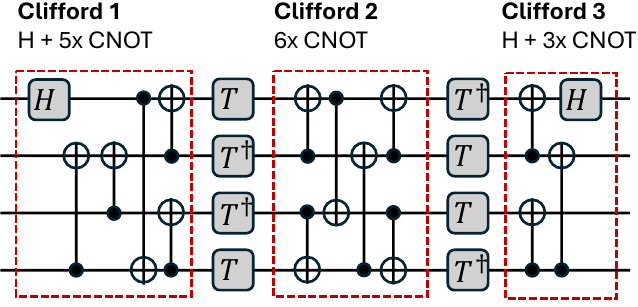}
    \caption{Toffoli Circuit Used for Simulation of Logical CliNR Protocol}
    \label{fig:toffoli_circuit}
\end{figure}

In this section we give details of the logical CliNR simulation. 
The two scenarios we considered were as follows:
\begin{enumerate}
    \item Four qubit Toffoli circuit: for this scenario we used a circuit from \cite{CNOT_MSD_synth} with the addition of a Hadamard conjugating the target logical qubit. This results in a circuit with three Clifford blocks composed of CNOT and Hadamard gates (see \Cref{fig:toffoli_circuit}).
    \item Random Clifford operators: \added{we used a set of 100 random Clifford operators} generated via the Qiskit  \texttt{random\_clifford\_tableau} function \cite{qiskit,random_clifford}.
    A circuit composed of single and two-qubit Clifford gates was generated from the Clifford using the Qiskit \texttt{synth\_clifford\_full} function.
\end{enumerate} 

We modeled the following protocols:
\begin{enumerate}
    \item Viterbi Gate-Based: for this option, we decomposed all single-qubit gates into two logical Pauli measurements and all two-qubit gates into three logical Pauli measurements to give a total number of logical measurements (LM in \Cref{tab:logical_CliNR}). 
    Each logical Pauli measurement is then performed via the Viterbi measurement protocol of \Cref{sec:meas_intro}, and the total number of cat-based measurements is summed to give CM in \Cref{tab:logical_CliNR}.
    \item  Viterbi CliNR: for this option, we used the CliNR protocol of \Cref{sec:logical_CliNR} to generate a set of logical stabilizer generators across auxiliary blocks A and B - this is the number of logical measurements. Each of the logical stabilizers was then measured using the Viterbi measurement protocol giving a total number of cat-based measurements (CM).
    \item MEDM CliNR: this uses the same logical stabilizer generators (and the same number of logical measurements) as the Viterbi CliNR method but instead measures these using the MEDM protocol of \Cref{sec:MEDM} to give the total number of cat-based measurements.
\end{enumerate} 
\added{For the CliNR protocol, we also calculated a low-weight representative for each stabilizer acting on auxiliary block B and verified that it was at most weight 18 for Q70 and 30 for Q102.}
Full results of the simulations are set out in \Cref{tab:logical_CliNR}. 
We see that implementing logical CliNR using Viterbi measurements of each of the stabilizers of the resource state requires between 2.2 and 2.6 times more cat-based measurements than implementing these using the MEDM protocol - this is consistent with the results of \Cref{tab:MEDM}.
We see a further gain when comparing the CliNR MEDM protocol with the Viterbi gate-based  protocol. 
This is because CliNR involves offline preparation of a resource stabilizer state which requires only $2\ell$ logical measurements where $\ell$ is the number of qubits in the support of the logical Clifford.
Note that this is the same as the bound in \cite{kliuchnikov2026clifford}.
The asymptotic bound on the number of 2-qubit gates required is $\mathcal{O}(\ell^2/\log \ell)$, each of which requires 3 logical measurements (see \cite{aaronson_improved_2004}). 
Even a circuit comprising $\ell$ single-qubit Clifford gates would require $2\ell$ logical measurements.
\begin{table}[hbt]
    \centering
    \begin{tabular}{|l|	r|	r|	r|	r|	r|	r|	r|	r|}
    \hline
&	\multicolumn{2}{c|}{\makecell[c]{Viterbi\\GB}}&	\multicolumn{2}{c|}{\makecell[c]{Viterbi\\CliNR}}&	\multicolumn{2}{c|}{\makecell[c]{MEDM\\CliNR}}&	\multicolumn{2}{c|}{\makecell[r]{MEDM CliNR\\Reduction}}\\
\hline
&	LM&	CM&	LM&	CM&	LM&	CM&	\makecell[c]{Viterbi\\GB}&	\makecell[c]{Viterbi\\CliNR}\\
\hline
\multicolumn{9}{|l|}{\textbf{1. Toffoli Circuit}}\\
\hline
Q70&	46&	232.9&	24&	121.5&	24&	54.9&	4.2x&	2.2x\\
Q102&	230&	1164.7&	120&	607.7&	120&	229.6&	5.1x&	2.6x\\
\hline
\multicolumn{9}{|l|}{\textbf{2. Random Clifford}}\\
\hline
Q70&	80&	406.0&	10&	50.6&	10&	21.9&	18.6x&	2.3x\\
Q102&	1158&	5865.6&	42&	212.7&	42&	80.2&	73.1x&	2.7x\\
\hline
    \end{tabular}
    \caption{Results for Logical CliNR Protocol: we model a 4-qubit logical Toffoli circuit which comprises 3 CNOT blocks and two transversal T blocks conjugated by Hadamard on one output qubit (see \Cref{fig:toffoli_circuit}). We used Q70 and Q102 code blocks from \cite{walking_cat}. As Q102 has $k=22$ logical qubits, we modeled 5 Toffoli circuits in parallel for this code so that $\ell=20$. The second scenario involves implementation of a \added{random set of 100 logical Cliffords} generated using the Qiskit \cite{qiskit} implementation of the algorithm in \cite{random_clifford}.
    As the Viterbi gate-based protocol requires an auxiliary logical qubit, we used random Cliffords on $\ell=k-1$ logical qubits.
    We show the number of logical Pauli measurements (LM) and the average number of cat-based measurements (CM) for three different protocols: the Viterbi gate-based protocol (Viterbi GB), the logical CliNR protocol where each stabilizer is measured individually using the Viterbi protocol (Viterbi CliNR) and the CliNR protocol where the stabilizers are measured using the multi-Pauli truncated error detected protocol (MEDM CliNR).
    In the final two columns, we calculate the ratio between the cat-based measurements required by Viterbi GB and Viterbi CliNR to MEDM CliNR.}
    \label{tab:logical_CliNR}
\end{table}

%

\end{document}